\documentclass[twocolumn,floatfix,prb,aps]{revtex4-1}
\usepackage{graphicx,float,placeins,url,color,nicefrac,color,multirow,amsmath,amssymb,amsfonts,tabularx,amsfonts}
\usepackage[dvipsnames]{xcolor}
\usepackage[unicode=true,colorlinks=true]{hyperref}
\hypersetup{linkcolor=blue,citecolor=blue,urlcolor=blue}
\begin{document}
\title{Post quench entropy growth in a chiral clock model}
\author{Naveen Nishad, M Santhosh, G J Sreejith}
\affiliation{Indian Institute of Science Education and Research, Pune 411008 India}
\date{\today}
\begin{abstract}
We numerically study quenches from a fully ordered state to the ferromagnetic regime of the chiral $\mathbb{Z}_3$ clock model, where the physics can be understood in terms of sparse domain walls of six flavors. 
As in the previously studied models, the ballistic spread of entangled domain wall pairs generated by the quench lead to a linear growth of entropy  with time, upto a time $\ell/2v_g$ in size-$\ell$ subsystems in the bulk where $v_g$ is the maximal group velocity of domain walls.
In small subsystems located in the bulk, the entropy continues to further grow towards $\ln 3$, as domain walls traverse the subsystem and increment the population of the two oppositely ordered states, restoring the $\mathbb{Z}_3$ symmetry.
The latter growth in entropy is seen also in small subsystems near an open boundary in a non-chiral clock model.
In contrast to this, in the case of the chiral model, the entropy of small subsystems near an open boundary saturates. We rationalize the difference in behavior in terms of qualitatively different scattering properties of domain walls at the open boundary in the chiral model.
We also present empirical results for entropy growth, correlation spread, and energies of longitudinal-field-induced bound states of domain wall pairs in the chiral model.
\end{abstract}
\maketitle

\section{Introduction\label{sec:intro}}
Quantum many-body dynamics in isolated systems has been an active area of contemporary research due in large part to realizations of tunable, almost isolated systems of long enough coherence times in cold atom experiments \cite{kinoshita2006quantum,hackermuller2010anomalous,trotzky2012probing,gring2011relaxation, Cheneau2012,ThermalCorrelationsEmergeBoson1D,langen2015experimental,MagnonBoundStates,Schreiber2015} A key notion in this context is the entanglement between the subsystem and environment. Though not as easily measurable in experiments\cite{Islam2015,GreinerEntanglementMeasurement} as local observables and correlation functions, entanglement in the eigenstates and its dynamics in general states give conceptual insights into broad questions of relaxation dynamics, dephasing, quantum measurements, thermalization, etc.\cite{PhysRevA.43.2046,PhysRevE.50.888,popescu2006entanglement,BardarsonMBLEntanglement,abanin2019colloquium,gogolin2016equilibration}. Entanglement is also relevant to practical considerations in quantum engineering and in designs of algorithms for quantum many-body dynamics. \cite{Schollwck2011}

A quench, in which an initial state with uncorrelated local observables undergoes a global change of the Hamiltonian, is a paradigmatic scenario that has been used to understand entanglement dynamics.\cite{essler2016quench} Under the new Hamiltonian, the initial state generically has a finite extensive energy density above the ground state. The initial state, which in general is not an eigenstate of the new Hamiltonian, evolves with time.
A large body of work on quenches in specific one dimensional systems has provided a semiclassical picture of the mechanism\cite{Calabrese2005} for the entanglement growth during this time evolution.\cite{Fagotti2008,Eisler2008,PhysRevB.90.205438,cotler2016entanglement,lauchli2008spreading,AlbaCalabrese2016,HuseKimBallisticEntanglement, Chiara2006,PhysRevB.94.214301,PhysRevB.102.094303,pomponio2019quasi}
Immediately after a quench, entangled quasiparticle pairs generated within short distances propagate away from each other. When these pairs are separated across the boundary between the subsystem and the environment, the subsystem effectively becomes entangled with its environment. The spreading of quasiparticles lead to decay of order parameters and induces correlations between initially uncorrelated local quantities in different parts of the system.\cite{PhysRevLett.106.227203,Calabrese2012a,Calabrese2012} Thus the quasiparticle dynamics is closely connected to the growth of entanglement and correlations. The entanglement growth in a subsystem of length $\ell$ is encoded in the following expression\cite{Calabrese2005}
\begin{multline}
S(t) \sim 2t\int_{v(k)<\frac{\ell}{2t}} dk v(k) f(k) + \ell \int_{v(k)>\frac{\ell}{2t}}dk f(k)
\end{multline}
where $v(k)$ represents the velocity of the quasiparticle indexed by quantum number $k$, and $f(k)$ is a function that depends on the amount of such quasiparticles produced at the time of the quench. If the dominant contribution to the integrals comes from a narrow range of $k$(with a velocity $v_m$), the first term produces a linear-in-time growth in entanglement till a time $\ell/2v_m$. At large times the second term dominates, and as the slowest quasiparticles cross the subsystem boundary, this term saturates to a constant proportional to $\ell$.

Studies on the 1D quantum transverse field Ising model (TFIM) and perturbations to this model have been crucial to guiding our intuition about post quench dynamics and relaxation in quantum chains.\cite{Calabrese2012a,PhysRevLett.106.227203,Calabrese2012,Kormos2017}. Ref-\onlinecite{Kormos2017} considered the non-equilibrium dynamics of a ferromagnetically ordered initial state, under a TFIM Hamiltonian with a longitudinal field perturbation aligned with the Ising order. Even a weak longitudinal field perturbation led to a strong suppression of the entanglement growth. The qualitative change in the entanglement dynamics could be attributed to the longitudinal field creating bound states of two different domain-wall-like quasiparticles of the Ising chain, preventing the original quasiparticles from spreading away from each other.

The TFIM is a $\mathbb{Z}_2$ symmetric member of a broad class of $\mathbb{Z}_n$ symmetric models with nearest neighbor interactions.\cite{Fendley2012} Simplest among these beyond the TFIM is the $\mathbb{Z}_3$ symmetric clock model. The $\mathbb{Z}_3$ model shares several features with the TFIM, such as a phase diagram with an ordered and paramagnetic phase and a continuous transition between them. The model can be transformed into a quadratic Hamiltonian of $\mathbb{Z}_3$ parafermions reminiscent of the quadratic Majorana Hamiltonian obtained following a Jordan Wigner transformation of the TFIM. The model generically has a chirality and has a richer set of domain wall flavors than the TFIM.

In this study, we numerically explore the dynamics after a weak quench in a $\mathbb{Z}_3$ symmetric chiral clock model with the goal of understanding the manner in which aspects of quench dynamics learned from TFIM extend to the $\mathbb{Z}_3$ chiral clock model, which has multiple domain wall flavors and chirality. Being non-integrable, we expect the clock model to thermalize \cite{Finch2018,rigol2008thermalization,polkovnikov2011colloquium}.
However, we will not focus on questions of long-time behavior and thermalization and instead explore the effect of chirality on entanglement growth at short times that can be reliably studied using numerical tools. 
We will work with weak quenches of a fully ordered state to a final Hamiltonian that is in a ferromagnetic regime of the model, where the low energy quasiparticles are long-lived domain walls. 
We will also explore the effect of the longitudinal field perturbations motivated by observations made in Ref-\onlinecite{Kormos2017}.
This being a numerical study, we focus on attributes easily accessible in the computational basis. The generation of entangled quasiparticle pairs can be pictured in the expansions in computational basis as generation of finite amplitudes, after quench, for states with flipped spin domains of various sizes centered around all points of the system. Domain walls flank these flipped spin domains. Dispersion of these domain walls and their scattering properties at the boundary will be used to understand the dynamics of the subsystem entropy.

The clock model is parametrized by a parameter $\theta$ that determines the chirality of the model, with  $\theta=0$ representing the non-chiral model.
After a quench, domain walls in the model, for any $\theta$, are produced in opposite chirality pairs (such as $...AAABBB..$ and $...BBBAAA...$), therefore opposite chirality domain walls are equally abundant. As we show, the domain walls propagate with a velocity independent of $\theta$ or the chirality of the domain walls. As a result, we find that qualitative features of the entanglement growth in the bulk of the system are same for the non-chiral and the chiral model.
Chirality however influences the scattering properties of domain walls at the open boundaries and hinders symmetry restoration in subsystems located close to the boundaries, preventing regions near the boundaries from thermalizing. The magnetization decays with time in the bulk of the system after a quench from the fully ordered state, indicating restoration of the $\mathbb{Z}_3$ symmetry in the final steady state. However, the magnetization at the boundary retains the initial value even in the steady-state. This can be related to a qualitatively different entanglement growth in small subsystems located in the bulk and at the boundary. Entanglement entropy of small subsystems in bulk continues to grow for times beyond the expected saturation time of $\ell/2v_g$ ($v_g$ being the maximal group velocity of the domain walls), whereas at the boundary, the entanglement entropy saturates after this time scale. The robustness of magnetization near the boundary can also be interpreted in terms of long coherence times near the boundary in systems with strong zero modes.\cite{Kemp_2017,PhysRevX.7.041062} Our work gives a complementary microscopic perspective for the same physics.

We describe the $\mathbb{Z}_3$ chiral clock model in Sec. \ref{sec:model}. For weak transverse fields, dynamics at low energies can be described in terms of far separated domain walls. Scattering properties of the domain walls at an open boundary are described in this limit. We will use this description to explain the contrasting behaviors of entropy growth in the small subsystems located near an open boundary of the system.
Section \ref{sec:numericalMethods} briefly describes the numerical time evolution calculations.
Results of the numerical simulations in the non-chiral and the chiral models are presented following this in Sec. {\ref{sec:numericalResults}} and Sec. \ref{sec:numResultsC} respectively.
We conclude with a summary of the results in Sec. \ref{sec:conclusions}. 
\section{Model}
\label{sec:model}
This study explores the growth of entanglement after a fully ordered initial state (all spins in the same direction) undergoes a weak quench to a Hamiltonian with finite transverse field and small non-zero chirality (as described further below).
Domain wall pairs are nucleated from every part of the chain after the quench. These domain walls propagate under the dynamics induced by the transverse field and lead to correlations between local properties of different parts of the chain.
Introducing chirality in the model modifies the dynamics by creating a difference between energies of different domain wall flavors and modifies the scattering properties of domain walls at an open boundary. We aim to explore how chirality affects the entropy growth, correlation spread and magnetization. 

Here we begin by describing the model. The $\mathbb{Z}_3$ chiral clock model in one dimension \cite{PhysRevB.24.398,PhysRevB.24.5180,HOWES1983169,Fendley2012} has the following Hamiltonian
\begin{equation}
H=-Je^{\imath\theta}\sum_i\sigma_i\sigma_{i+1}^\dagger-fe^{\imath\phi}\sum_i\tau_i+\text{h.c.}\label{eq:bareChiralHamiltonian}
\end{equation}
where operators $\sigma_i$ and $\tau_i$ located at the $i^{\rm th}$ site are
\begin{equation}
\sigma=\begin{pmatrix}1 &0&0\\0&\omega&0\\0&0&\bar{\omega}\end{pmatrix}\;\;\tau=\begin{pmatrix}0&1&0\\0&0&1\\1&0&0\end{pmatrix}.
\end{equation}
Here $\omega={\rm exp}(2\pi\imath/3)$ and $\bar{\omega}={\rm exp}(-2\pi\imath/3)$. The algebra satisfied by the above operators: $\sigma_i^3=\tau_i^3=1$  and $\sigma_i\tau_j=\delta_{ij}{\bar{\omega}}\tau_j\sigma_i$ presents a $\mathbb{Z}_3$ analogue of the algebra of Pauli matrices $\sigma_z$ and $\sigma_x$; and the Hamiltonian forms a $\mathbb{Z}_3$ symmetric analogue of the $\mathbb{Z}_2$ symmetric spin-$\frac{1}{2}$ transverse field Ising model.\cite{PFEUTY197079} 
The Hamiltonian commutes with the $\mathbb{Z}_3$ generalization of the parity operator namely $P=\prod \tau_i$, which allows labeling of energy eigenstates with parity eigenvalues $1,\omega$, or $\bar{\omega}$. 
For simplicity we will work with systems with $\phi=0$ and will use units where $J=\hbar=1$. 
The chirality of the model is determined by $\theta$ and can be assumed to take values in the range $[0,2\pi/3]$, as the physics at $\theta$ can be related to $\theta+2\pi/3$ through a local unitary transformation by $\prod_i \tau_i^{i}$.

In the absence of a transverse field ($f=0$), energy eigenstates are direct products of $\sigma_i$ eigenstates at each site with energies $ -2J\sum_i\cos(\theta + \alpha_{i,i+1})$ where $\alpha_{i,i+1} \in \{0,\pm 2\pi/3\}$ is $\arg (\langle\sigma_{i}\rangle/\langle\sigma_{i+1}\rangle)$. For $\theta$ in $[0,\pi/3]$, the ground state is described by $\alpha=0$, corresponding to all spins pointing in the same direction ($1$, $\omega$ or $\bar{\omega}$). The simplest excitations are localized domain walls. In the non-chiral model the opposite chirality domain walls ($...AABB...$ and $...BBAA...$) as well as domain walls at different locations are degenerate. The ground state is ferromagnetic in the entire range $[0,\pi/3]$ but finite $\theta$ causes an energy difference ${2}J\sqrt{3}\sin \theta$ between domain walls of opposite chirality. The ground states in the regime $\theta>\pi/3$ have a twisted ordering with adjacent spins $\langle \sigma_i \rangle$ differing by a factor of $\omega$. These domain walls disperse if the transverse field is non-zero, lifting the degeneracy of different domain wall states. 

We will consider quenches to Hamiltonians with finite $f$, and a non-zero $\theta\in {(}0,\pi/6]$, {\emph {i.e.}} in the regime where the classical ground state is still ferromagnetic but $\theta$ influences the dynamics by inducing a chirality to the domain walls. We find that quenches to larger $\theta$ result in more complex domain wall dynamics due to the possibility of a domain wall splitting into two as discussed later in this section.

The non-chiral model ($\theta=0$) is ferromagnetic for $f<J$ and exhibits a continuous phase transition to a paramagnetic phase ($f>J$).\cite{Fendley2012,Motruk2013} 
The ground state in the ferromagnetic phase is three-fold degenerate (forming a parity multiplet) - with a splitting that decays exponentially with system size; but the excited states (for $f\neq 0$) have parity multiplets that show a power law decay of the splitting with system size.\cite{Fendley2012,Fendley2014} 
The chiral model with finite $\theta$ also shows a transition from a $\mathbb{Z}_3$ symmetry broken phase to the paramagnetic phase at some $f_c\lesssim J$.\cite{Zhuang2015,RhineNumerical,PhysRevB.97.014309} Unlike the non-chiral case, the excited states in the  broken $\mathbb{Z}_3$ symmetry phase have multiplets with a splitting that exponentially decays with system size.
For weak transverse fields ({$2f<J\sqrt{3}\sin\theta$}), this degeneracy can be attributed to a weak zero-energy parafermion mode localized at the boundary in the Jordan Wigner transformed dual model.\cite{Fendley2012} 

The power law decay with system size of the splitting in the non-chiral model can be understood as arising from the scattering properties at the boundary.\cite{Fendley2014} We present a simplified form of this model here and will use this as a basis to rationalize the entropy growth in subsystems near the boundary. In the limit of low energy densities, the states in the model can be understood in terms of a dilute set of domain walls, and interaction between domain walls may be neglected. For the discussions below, we will assume that transitions to the zero-domain-wall and two-domain-wall states are suppressed by an energy gap. 

Now we focus on the dynamics of a single domain wall in the vicinity of a boundary. We denote by $|AB,i\rangle$ a direct product state representing a domain wall on the bond $i$ separating regions of $\langle \sigma \rangle=A$ to the left and $\langle \sigma \rangle=B$ to the right. There are six possible domain wall types, but an incoming domain wall, say of type $1\omega$ (with $\langle \sigma \rangle = 1$ to the left of $i$ all the way to $-\infty$) that approaches a boundary on the right, can be reflected only as a $1\omega$ or $1\bar{\omega}$ domain wall types at the boundary. Transition to any of the other domain walls such as $\omega1$ will require global changes in the spin states.

\begin{figure}
\includegraphics[width=0.85\columnwidth]{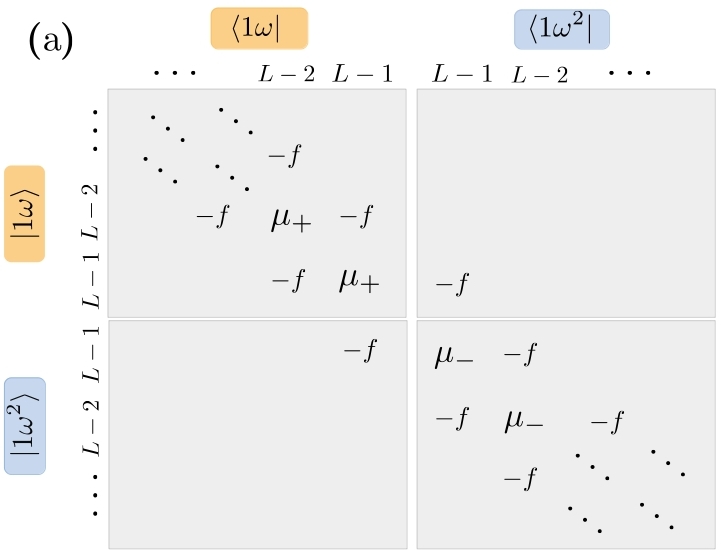}
\includegraphics[width=\columnwidth]{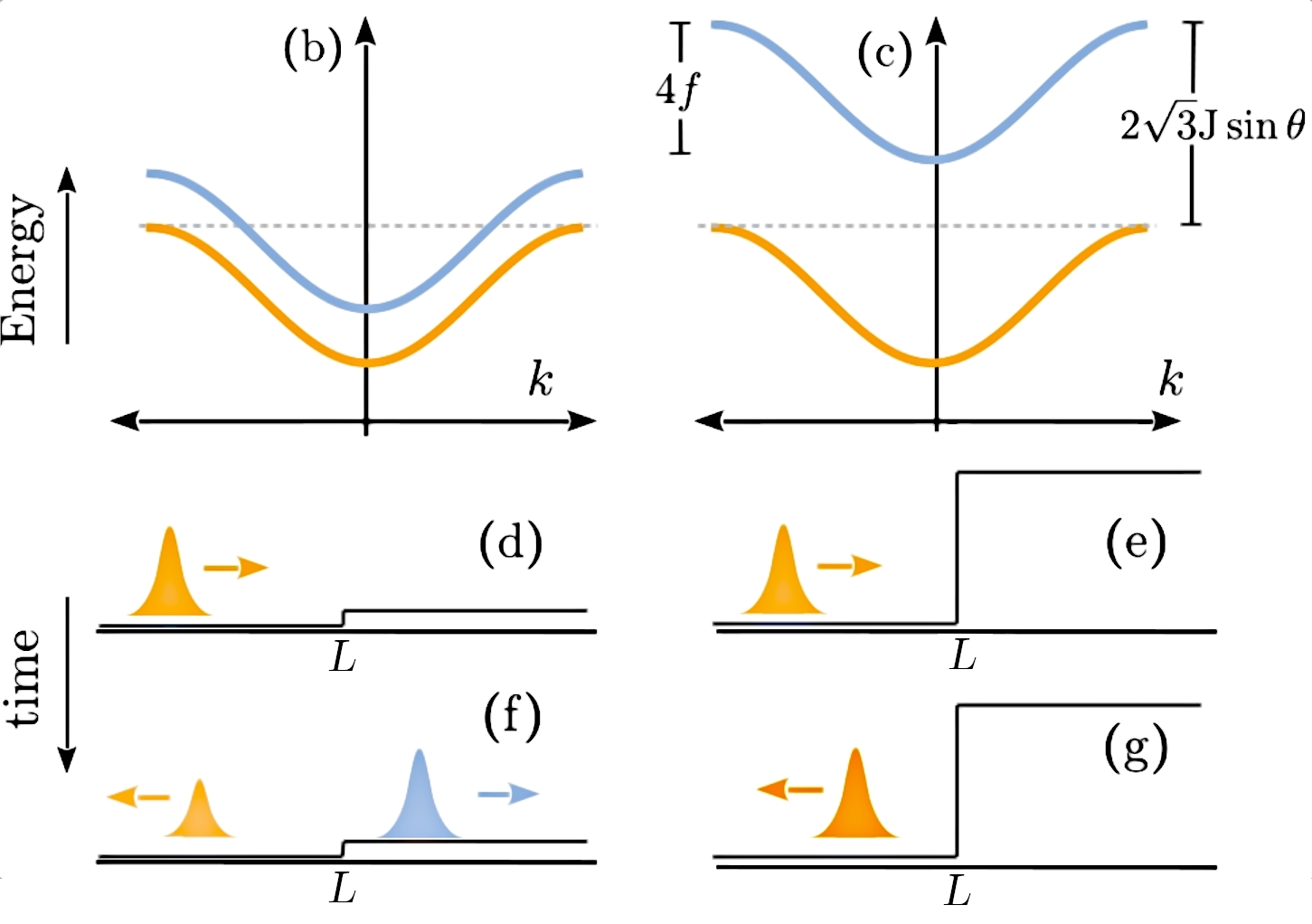}
\caption{Panel a shows the Hamiltonian matrix in the one-domain-wall space after relabeling $1\bar{\omega}$ domain wall at bond number $i<L$ as $x=2L-i-1$ and $1\omega$ domain wall at bond number $i<L$ as $x=i$. Effective dispersions of opposite chirality (the two colors represent the bands for the two different chiralities) domain walls are shown schematically for the case of small $\theta$(panel b) and large $\theta$(panel c). Panels d and f schematically show the fate of a domain wall wave-packet that bounces off a boundary in a system with small $\theta$. Panel d shows the incoming packet and panel f shows the fate after collision with the boundary. Incoming domain wall has one chirality (indicated in orange) whereas the reflected domain wall is primarily of opposite (blue) chirality.
Panels e and g are similar but for a case where $\theta$ is larger. Here the domain wall bounces back without change in the flavor.  \label{fig:BoundaryReflection}}
\end{figure}

The Hamiltonian projected into the space of these states can be written as 
\begin{equation}
PHP = H_{1\omega} + H_{1\bar{\omega}} + H_{\rm boundary}.
\end{equation}
Here the Hamiltonian for each flavor of the domain wall is given by 
\begin{eqnarray}
H_{1\omega} = \mu_+\sum_i|1\omega,i\rangle\langle1\omega,i| - f\sum_i|1\omega,i\rangle\langle1\omega,i+1|+{\rm h.c.}\nonumber \\
H_{1{\bar \omega}} = \mu_-\sum_i|1{\bar \omega},i\rangle\langle1{\bar \omega},i| - f\sum_i|1{\bar \omega},i\rangle\langle1{\bar \omega},i+1|+{\rm h.c.}\nonumber
\end{eqnarray}
where 
\begin{equation}
\mu_{\pm} = 2J [ \cos\theta - \cos(\theta \mp 2\pi/3) ]\label{eq:domainwallenergy}
\end{equation}
Away from the boundaries, and away from each other, the Hamiltonian imparts a dispersion of $\epsilon^\pm_k = \mu_\pm -2f\cos(k)$ to the domain walls. 
The boundary scatters between the two relevant domain wall types:
\begin{equation}
H_{\rm boundary}=-f|1\omega,L-1\rangle\langle1\bar{\omega},L-1| + {\rm h.c.}\nonumber
\end{equation}
The Hamiltonian is a tridiagonal matrix shown in Fig. \ref{fig:BoundaryReflection}(a). Relabeling the basis states as $\left\Vert x=i \right \rangle = \left | 1\omega,i \right \rangle$ and $\left|| x=2L-i-1\right \rangle = \left | 1{\bar \omega},i \right \rangle$, this represents the Hamiltonian of a particle with kinetic energy $-2f\cos(k)$ traveling across a potential jump $\Delta = \mu_+-\mu_-=2\sqrt{3}J\sin\theta$ at the position $L$. In the relabeled form, states propagating away on the right of $L$ represent, physically, a $1\bar{\omega}$ domain wall reflecting back towards the left from the boundary. For $\theta=0$, gap $\Delta$ is zero and the particle tunnels across with unit probability; {\emph {i.e.}} there is a complete reflection of the $1\omega$ to a $1\bar{\omega}$ domain wall.\cite{Chim1995} For larger $\theta$ such that the bandwidth is smaller than the gap, {\emph {i.e.}} $2\sqrt{3}J\sin\theta>4f$ the domain wall bounces back without any change in its flavor.

Boundary mediated tunneling from one domain wall flavor to another results in an increased energy splitting in excited states of the non-chiral model.\cite{Fendley2014} The nature of the zero mode and the analysis of the energy splitting are not directly related to the present work, but we will use the above effective model to make sense of the numerical results.

As $\theta$ approaches $\pi/6$, Eqn. \ref{eq:domainwallenergy} suggests that $\mu_+\sim \mu_-/2$; so the energy of a domain wall of the form $1\bar{\omega}$ is same as that of a pair of domain walls of opposite chirality $1{\omega}$ and $\omega\bar{\omega}$. Thus the $1\bar{\omega}$ can evolve into a domain wall pair of the form $1\omega\bar{\omega}$. We restrict to a discussion of the regime where the domain wall is stable. In the rest of this manuscript, we describe the results from numerical simulations of the quenches in the limit of small $f$, $\theta<\pi/6$, and $\phi=0$ regime of the clock model.

\section{Numerical simulation of the time evolution}
\label{sec:numericalMethods}
States and operators are represented as matrix product states and matrix product operators \cite{Schollwock2011} respectively, with a maximum bond dimension of 300. Time evolution of the states were implemented by using fourth order Suzuki-Trotter approximant \cite{Hatano2005} to represent exp$(-\imath H\delta t)$ with time steps $\delta t=10^{-3}$. This approximant decomposes the unitary operator as a sequence of two site gates acting on adjacent sites. Further details of the numerical implementation for the approximant is similar to that used in Ref \onlinecite{Naveen2020} and has been summarized in the Appendix.
\begin{figure}
\includegraphics[width=\columnwidth]{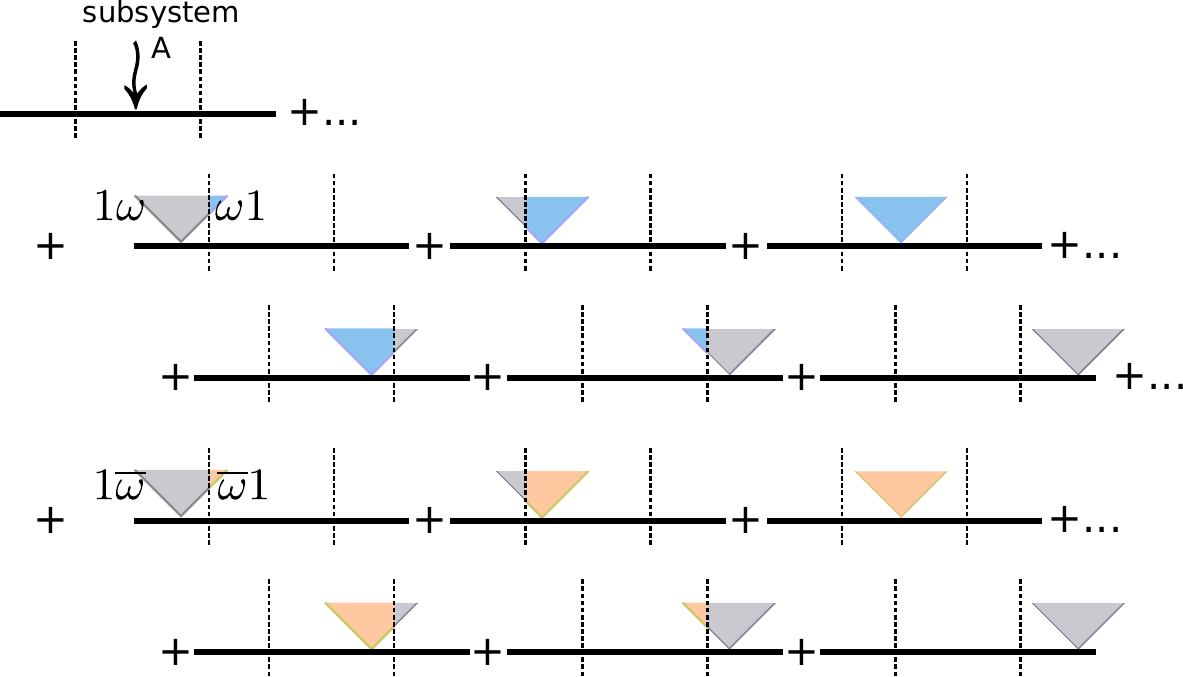}
\caption{
Schematic representation of the state of the system after a weak quench. State after the quench is, to a good approximation, a linear combination of the ordered state and $1\omega1$ and $1\bar{\omega}1$ type two domain wall states. The vertical direction represents the time evolution of domains in each term. Flipped spin domains are nucleated from all parts of the chain forming different terms of the linear combination in the computational basis. The domains expand as the domain walls propagate. In a system with non-linear dispersion, the domain walls of different momenta propagate at different group velocities. Dashed vertical lines demarcate a subsystem. For each cone, spins flipped to $\omega$ and $\bar{\omega}$ inside the subsystem are shown in blue and orange colors. The flipped spins outside the system are colored gray irrespective of direction of the spins inside them.
\label{fig:linearCombSchematic}}
\end{figure}

\label{sec:numResultsNC}

\begin{figure}
\includegraphics[width=\columnwidth]{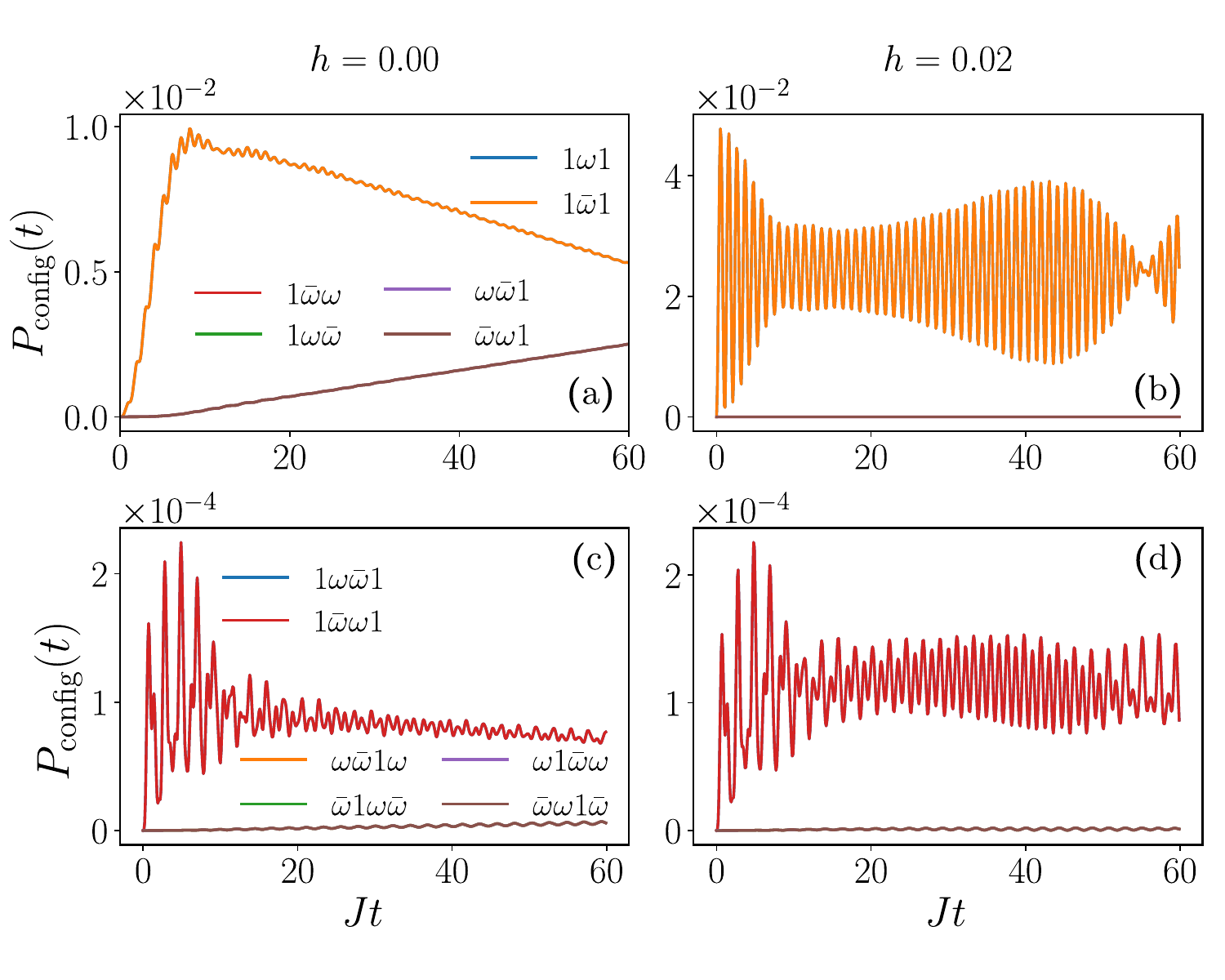}
\caption{
Panel a shows the total probability of domain wall pair states that occur with non-zero probabilities after the quench from a fully ordered state to a final Hamiltonian with $f=0.1$ and $\theta=0$. System size is $L=40$. Panel b shows the probabilities but in a system with an additional longitudinal field $h=0.02$ in the final Hamiltonian. 
In Panel a and b, domain wall types $1\omega 1$ and $1\bar{\omega}1$ have equal probabilities and the corresponding lines (orange and blue) completely overlay over one another. Similarly, lines corresponding to domain wall configurations $1\omega\bar{\omega}$, $\bar{\omega}\omega 1$, $1\bar{\omega}\omega$ and $\omega\bar{\omega}1$ also overlay over each other.
Panels (c,d) show the corresponding results for the three domain wall states of the form $ABCA$.
\label{fig:NCDomainWallProb}
}
\end{figure}
\section{Numerical results: Quench into the non-chiral model}
\label{sec:numericalResults}

In this section, we describe the dynamics after an initial state $\psi_0$, in which all sites are in the $\langle\sigma\rangle=1$ direction, is quenched to the non-chiral Hamiltonian at finite transverse field. After the quench, the system evolves into a linear combination of the initial state and (with small amplitudes) domain wall pair states  of the form $1\omega 1$ and $1\bar{\omega}1$. The flipped spin domains are nucleated from every part of the chain, and domain walls on the opposite sides of the flipped spin domain propagate in opposite directions with a characteristic velocity corresponding to the maximal group velocity $v_g=2f$ (Sec. \ref{sec:model}) of the domain walls; thereby expanding the flipped spin domains. This is schematically represented in Fig \ref{fig:linearCombSchematic}.

Figure \ref{fig:NCDomainWallProb}a presents the total probability weight (over all positions) of two domain wall states of different types, showing that among these, the $1\omega 1$ and $1\bar{\omega}1$ are equally populated. Small domains of size $1$ (where domain walls are separated by a distance of 1 lattice unit) show rapid oscillations and have been omitted. 
As time progresses, the domain wall pair states of the form $1\omega1$ ($1\bar{\omega}1$) formed in the vicinity of the left-hand-side boundary reflect off the boundary as a $\bar{\omega}\omega1$ ($\omega\bar{\omega}1$) domain wall pair states (as described in Sec \ref{sec:model}). On the right hand side boundary, the domain walls scatter from the $1\omega1$ ($1\bar{\omega}1$) state into $1\omega \bar{\omega}$ ($1\bar{\omega}\omega$) state. Since the domain walls reach the boundary with a characteristic rate $v_g$, there is a linear rate of decrease of the population of the $1\omega1$ and $1\bar{\omega}1$ states as shown in Fig \ref{fig:NCDomainWallProb}a. 
Correspondingly the population of the states of the types $1\omega\bar{\omega}$, $1\bar{\omega}\omega$, and $\omega\bar{\omega}1$, $\bar{\omega}\omega 1$ linearly increase with time. 

In the presence of an additional longitudinal field in the final Hamiltonian,
\begin{equation}
H_{\rm longitudinal} = -h(\sigma + \sigma^\dagger)
\end{equation}
the energy of the flipped spin domains have an (positive) energy contribution that grows linearly with the domain size. The domain wall pairs now appear to attract with an energy linear in the distance between them \cite{Kormos2017}. With this constrained domain wall dynamics, scattering processes at the boundary are suppressed as indicated by a constant probability on an average of the $1\omega1$ and $1\bar{\omega}1$ states in Fig \ref{fig:NCDomainWallProb}b.
(as opposed to a linear decay in the absence of $h$).

\begin{figure}
\includegraphics[width=\columnwidth]{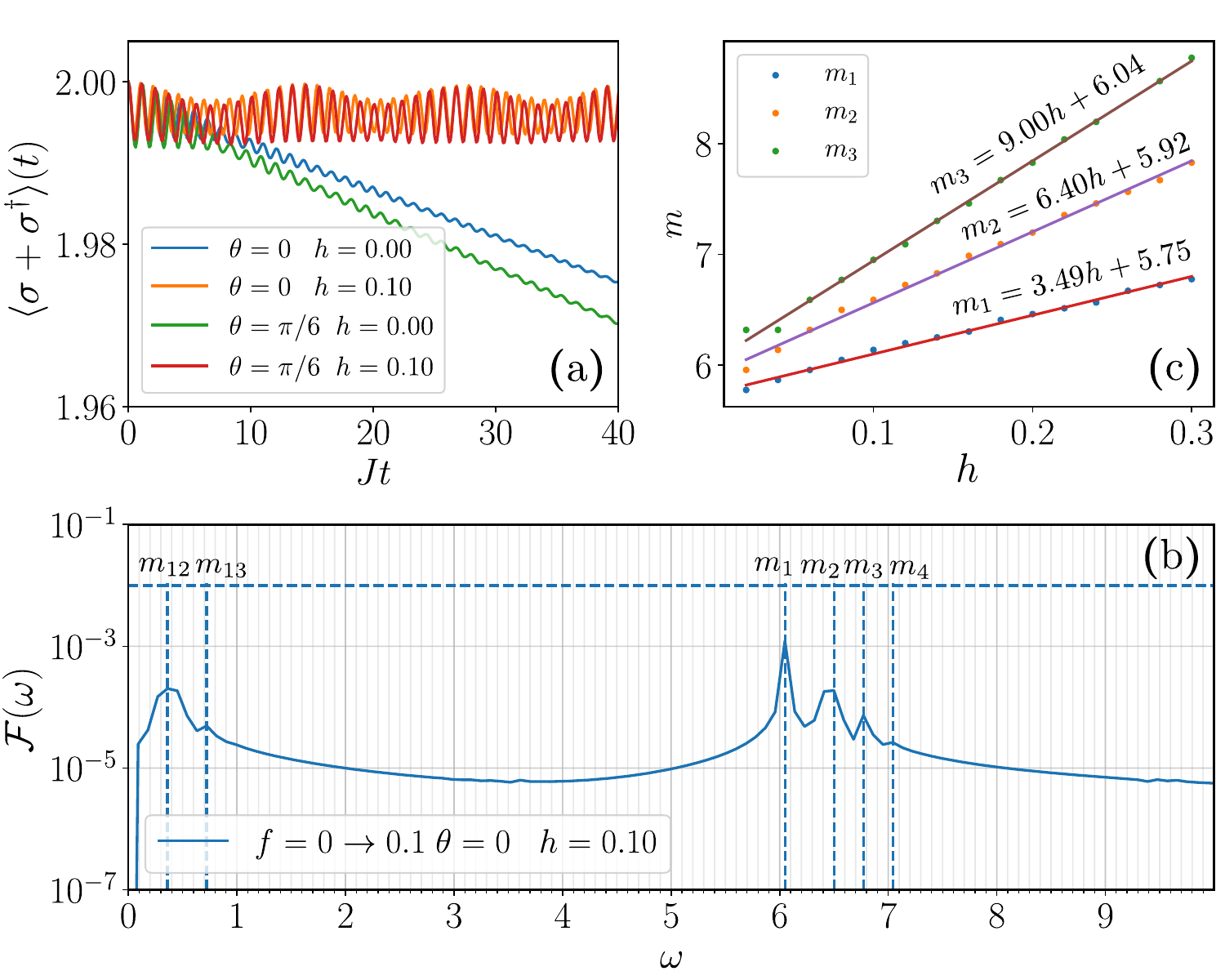}
\caption{Magnetization as a function of time for quenches to different final Hamiltonians is shown in panel a. Panel b shows the power spectrum of magnetization for a specific example where the final non-chiral Hamiltonian has $f=0.1,h=0.1$. The peaks correspond to the masses $m_i$ of the domain wall bound states or the differences between the masses $m_{ij}=|m_i-m_j|$. Variation of first three masses with longitudinal field $h$ is shown in panel c. System size used is $L=40$.
\label{fig:Mag_NC}}
\end{figure}

\subsection{Magnetization}
Here we present the results regarding local magnetization in the bulk of the system. Magnetization $\langle M \rangle =\langle\sigma+\sigma^\dagger\rangle/2$ is $1$ in the initial state. After a time $t$ from the quench, domain walls that originate within a neighborhood of radius $\sim  v_gt$ around a site $i$ cross this site at time $t$ thereby reducing the local population of the state $1$ at the site and increasing population of $\omega$ or $\bar{\omega}$; and decreasing the local magnetization at $i$ linearly with time as shown in Fig \ref{fig:Mag_NC}a.

The instantaneous magnetization can be expressed in the eigenbasis of the Hamiltonian as
\begin{equation}
\langle M \rangle = \langle \psi(t) | M | \psi(t) \rangle = \sum_{i,j} \bar{c}_i c_j e^{\imath t(E_i-E_j)} M_{ij}.
\end{equation}
Here $c_i$ are the coefficients in the expansion of the initial state in the eigenbasis of the Hamiltonian and $M_{ij}$ is the matrix element of a local magnetization in the eigenbasis of the Hamiltonian. This indicates that the power spectrum of the time dependent oscillations of the magnetization carries the information of the gaps between finitely populated energy eigenstates. Peaks in this power spectrum occur at frequencies equal to the gaps between parts of the energy spectrum with a large energy-density-of-states (such as the bottom of the domain wall dispersion) or eigenstates with a large population (such as the ground state). Consistent with this, we find that the oscillatory part of the magnetization has a frequency peak equal to the gap between the ground state and the minimal kinetic energy of domain wall pairs:
\begin{equation}
m_2(\theta=0) = \epsilon_{k=0}^+ + \epsilon_{k=0}^- = 6J - 4f
\end{equation}
In the presence of a longitudinal field, confinement of domain wall pairs prevents decrease in magnetization as shown in Fig \ref{fig:Mag_NC}a. The attractive interaction results in bound states of domain wall pairs. The energy minima of the dispersion of bound domain-wall-pairs (or equivalently the masses of bound domain wall pairs) can be extracted from the spectral peaks in the oscillatory part of the magnetization. The power spectrum of the magnetization oscillations at a finite longitudinal field is shown in Fig \ref{fig:Mag_NC}b. The peaks depend on $h$. A set of peaks split off from the one at $m_2(\theta=0)$ as $h$ is increased from $0$ to finite values; these frequencies are labeled $m_1,m_2,m_3\dots$ and can be associated with the masses of different domain wall bound pairs. The frequencies of the peaks located at the lower end of the power spectrum match with the differences between these masses. Figure \ref{fig:Mag_NC}c shows the variation of the bound pair energies as a function of the longitudinal field $h$.

\begin{figure}
\includegraphics[width=\columnwidth]{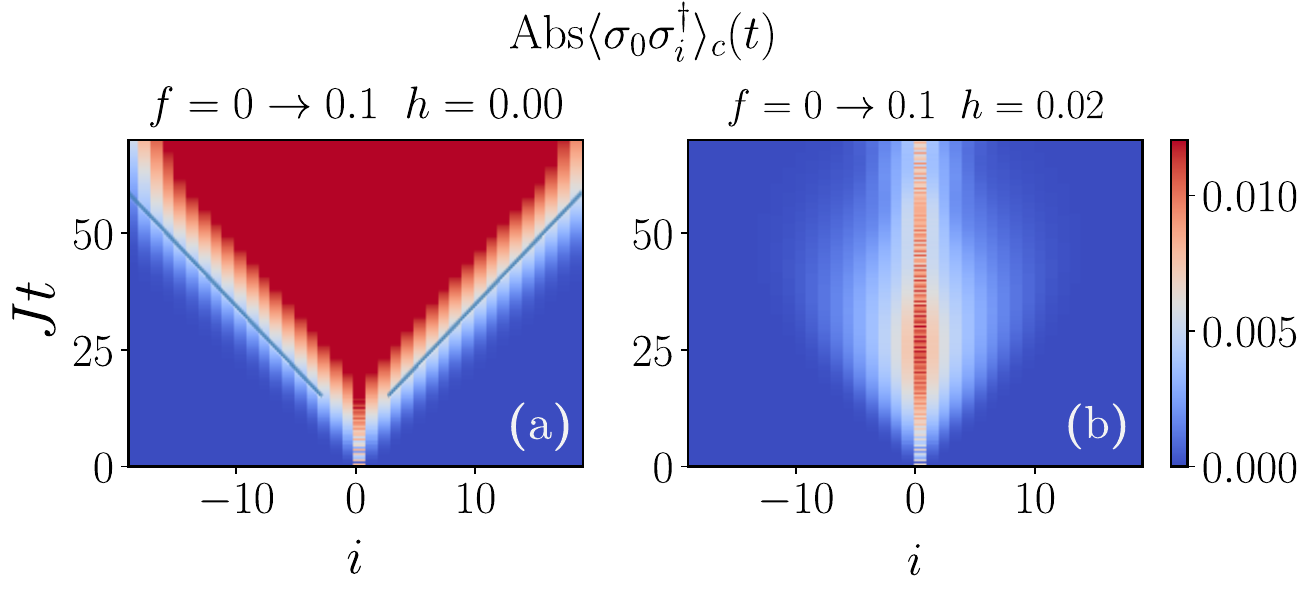}
\caption{Panel a shows the absolute value of the connected correlations as a function of position and time in the case of a quench to the non-chiral clock model. The lines show constant height contours and have a slope of $~1.0/0.38$ consistent with the expected correlation spread rate of $2v_g=4f\sim0.4$.
Panel b shows same quantity in the case where the final Hamiltonian has an additional longitudinal field constraining the spread of domains. System size used is $L=40$.
\label{fig:correlations_nc}}
\end{figure}

\begin{figure*}
\includegraphics[width=\textwidth]{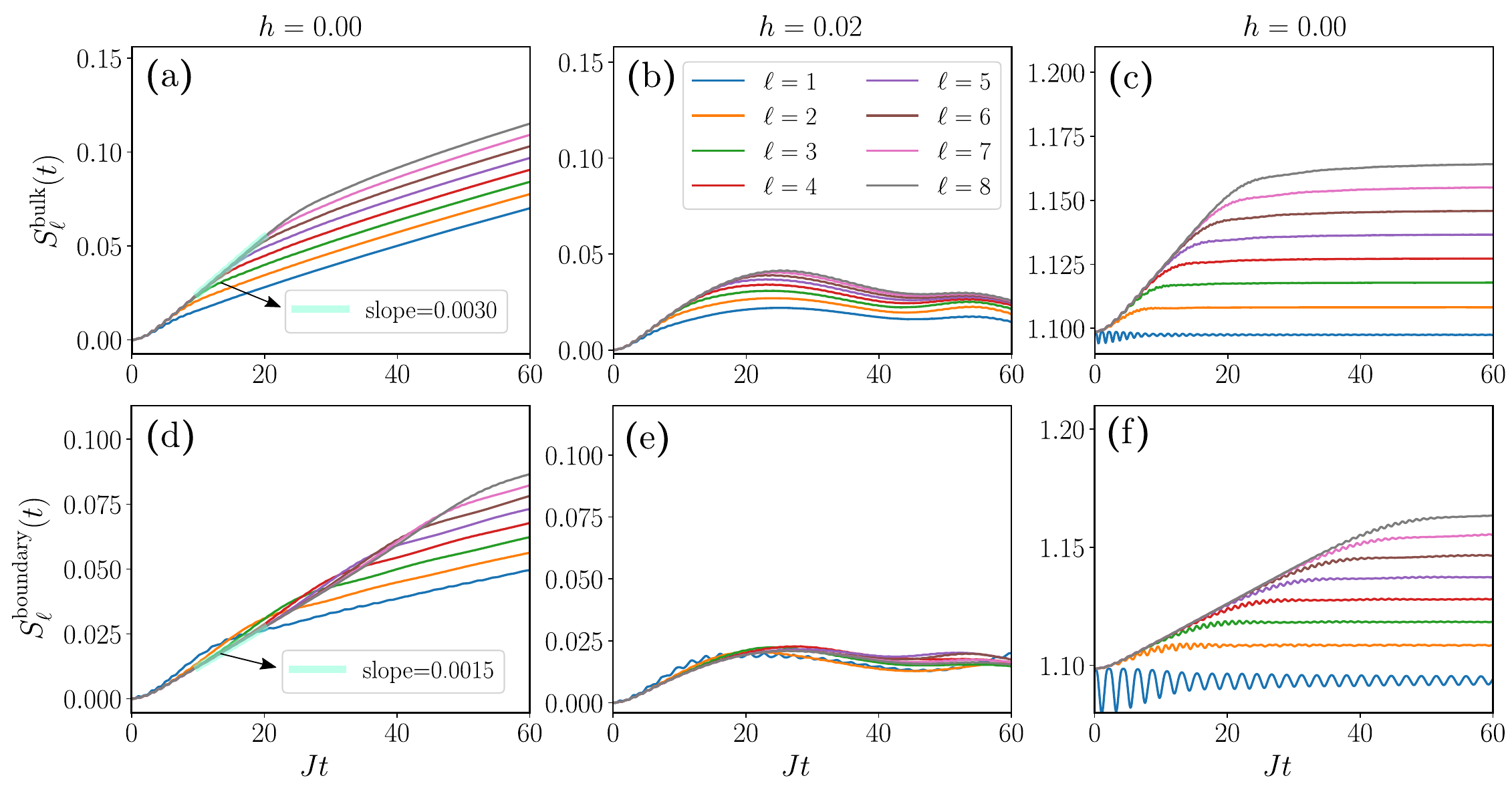}
\caption{
Panel a shows entanglement growth in a subsystem in the bulk after a quench to the non-chiral Hamiltonian from a fully ordered initial state. The results for a subsystem at the boundary of the system is shown in panel d. Panels b and e show the corresponding results for the case of a quench to a Hamiltonian with an additional longitudinal field, demonstrating the suppression of entanglement growth. Panels c and f show the entropy growth in the bulk and near the boundary in a scenario where the initial state is the parity eigenstate  $|11...1\rangle+|\omega\omega ...\omega\rangle+|\bar\omega\bar\omega ...\bar\omega\rangle$. System size simulated is $L=40$.
\label{fig:NC_Entropy}}
\end{figure*}

\subsection{Two point correlations}
We now consider the connected, equal time, correlations between local operators at spatially separated pair of points in the bulk. In particular we focus on $C(r,t)=\langle \psi(t)| \sigma_0 \sigma^\dagger _r| \psi(t)\rangle_c $. We expect qualitative features of spread of correlations between other generic local operators to be the same; we focus on this as its imaginary component shows a non-zero (zero) value in a quench to the chiral (non-chiral) Hamiltonian.

The correlation $C$ is zero everywhere in the initial state. The correlation $C$ expanded in the computational basis shows that $C(r,t)$ is non zero if there are flipped spin domains that extend from $0$ to $r$. The first among such domains appear when the domain wall pairs nucleated from $r/2$ at time $0$ reach positions $0$ and $r$ at time $t=r/2v_g$. As a result the correlations $C(r,t)$ spread with a velocity $2v_g\sim4f$. Correlation functions plotted in Fig \ref{fig:correlations_nc}a show the linearly expanding region with finite correlations. As expected from confinement of domain wall pairs, the presence of the longitudinal field suppresses the spread of correlations (Fig \ref{fig:correlations_nc}b). In all cases we find that the imaginary part of the correlation is zero (this is guaranteed by translation symmetry of the initial state and final Hamiltonian in the bulk, and spatial parity symmetry).

\subsection{Entanglement entropy}
In this section, we present the numerical results for entanglement entropy growth in small subsystems after the system initially in the fully ordered state ($...1111...$) is quenched to a non-chiral Hamiltonian at finite $f$. The subsystems are initially unentangled. Shortly after the quench, the time evolved state is a linear combination of the fully ordered initial state and, with small amplitudes, states with flipped spin domains of typical size $\sim 2v_g t$ that were nucleated from every part of the chain at time $t=0$ (Fig \ref{fig:linearCombSchematic}).

In order to evaluate the reduced density matrix of a contiguous segment $A$ of size $\ell$, the complementary region is traced out. Initially the subsystem is in a pure state with only the fully ordered state $|...111...\rangle_A$ populated. The entanglement entropy increases with time as progressively more flipped spin domains nucleated near the boundary of $A$ (on either side of the boundary) cross the boundary. As time progresses more states of the form $|\omega\omega...\omega11...1\rangle_A$, $|\bar{\omega}\bar{\omega}..\bar{\omega}11...1\rangle_A$, $|1...11\omega\omega..\omega\rangle_A$, and $|1...11\bar{\omega}\bar{\omega}..\bar{\omega}\rangle_A$ are populated.
Number of such states that are populated grow linearly with time initially. This results in a growth of entropy that is linear in time. Fig \ref{fig:NC_Entropy}a shows entropy as a function of time in a small subsystem in the bulk. A rough estimate of the entanglement growth rate can be obtained from the data presented in Fig \ref{fig:NCDomainWallProb}a. Probability $p$ associated with domain walls nucleated from each point in space can be estimated to be $1/L$ fraction of the total domain wall probabilities. As domain walls propagate into the subsystem previously unpopulated state of the subsystem is populated with a probability weight $p$. This adds an entropy of $s=-p\ln p$. Counting two kinds of domains ($1\omega1$ and $1\bar{\omega}1$) crossing the two boundaries in either directions at a typical rate $\sim v_g$, the entropy growth rate is $\lambda=8v_gs$. From Fig \ref{fig:NCDomainWallProb}, $p\approx 0.009/40$, resulting in $\lambda=0.003$ which is close to the numerically obtained value in Fig \ref{fig:NC_Entropy}a. This estimate ignores that the group velocity is not the same for all domain wall momenta and that there are off-diagonal entries in the density matrix.

At time $t=\ell/2v_g$, the domain wall pairs that originated in the {vicinity} of the center of $A$ exit the subsystem. For $t>\ell/2v_g$, this equals the number of new domain walls that enter the system, resulting in a saturation of this mechanism of entanglement growth at an entropy value that is proportional to $\ell$. 

In a system that is initially prepared in the fully ordered $...1111...$ state, the exit of domain wall pairs that commence at $t=\ell/2v_g$ results in conversion of a fraction of the initial state $|..1111...\rangle_A$ into the oppositely ordered states $|...\omega\omega\omega...\rangle_A$ or $|...\bar{\omega}\bar{\omega}\bar{\omega}...\rangle_A$. Populations of these two oppositely ordered states increase with time as more and more domain walls exit the subsystem. This results in a further increase in the entropy after the expected saturation time of $\ell/2v_g$. We expect that the entropy of the small subsystem grows into that of a mixed state of all three ordered states with an entropy of $\sim \ln 3$. For large systems and for larger $f$, where the saturation entanglement is much larger than $\ln 3$, the latter growth will only provide a subleading contribution to total entanglement. 

The saturation of the initial mechanism of entanglement growth at a time $\ell/2v_g$ (where the maximal group velocity $v_g$ is $\sim2f$) as well as further growth of entanglement can be seen in Fig \ref{fig:NC_Entropy}a. Approach to $\ln 3$ is, unfortunately, not verifiable within the timescales of the simulations. As expected, entanglement growth is strongly suppressed even in the presence of a small longitudinal field (Fig \ref{fig:NC_Entropy}b). 

In contrast, entropy after a quench from an initial system prepared in one of the three fully ordered parity eigenstates starts from $\ln 3$ and increases with time linearly until the entropy saturates at the time $t=\ell/2v_g$. The above mentioned process which converts the population of $|...11111...\rangle_A$ into the oppositely ordered state in $A$ is compensated by the reverse process resulting in no growth of entanglement after a time $t=\ell/2v_g$. This can be seen in simulations of the entropy growth after quench from the initial state $|...11...\rangle+|...\omega\omega...\rangle+|...\bar{\omega}\bar{\omega}\bar{\omega}...\rangle$, presented in Fig \ref{fig:NC_Entropy}c.

\begin{figure}
\includegraphics[width=0.8\columnwidth]{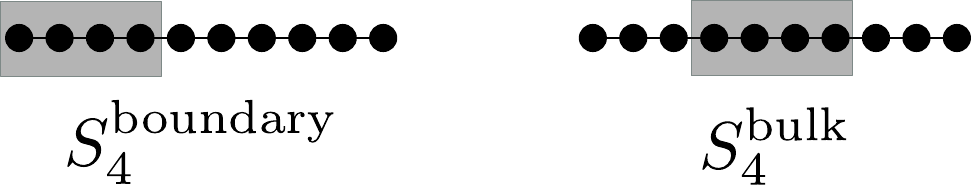}
\caption{
Illustration of an $\ell=4$ site subsystem located at the boundary (a) and one located in the bulk of the system (b). The entanglement entropy of these subsystems with the rest of the system are labeled as $S_\ell^{\rm boundary}$ and $S_\ell^{\rm bulk}$.
\label{fig:bulkOrBoundary}}
\end{figure}

For a subsystem located in the bulk, the entanglement growth occurs due to all domain walls that cross either one of the two boundaries of the subsystem. A subsystem located at the boundary of a system (Fig \ref{fig:bulkOrBoundary}) on the other hand shows entanglement growth at half the rate as domain walls  cross only one boundary. The saturation of this entropy growth occurs at a time when a domain wall pair nucleated at the boundary of the system exits the subsystem. This happens at the time $t=\ell/v_g$ when the domain wall pairs nucleated at the edge of the system at $t\sim0$ reach the inner boundary of the subsystem. Fig \ref{fig:NC_Entropy}d shows the entropy growth in a subsystem near the boundary for the same quench as in Fig.  \ref{fig:NC_Entropy}a. As expected, the entanglement growth rate at the boundary (Fig \ref{fig:NC_Entropy}d) is half of that in the bulk (Fig \ref{fig:NC_Entropy}a) and saturates in twice the time. Similar results hold in the case of parity eigenstate (Fig \ref{fig:NC_Entropy}f).

\begin{figure}
\includegraphics[width=\columnwidth]{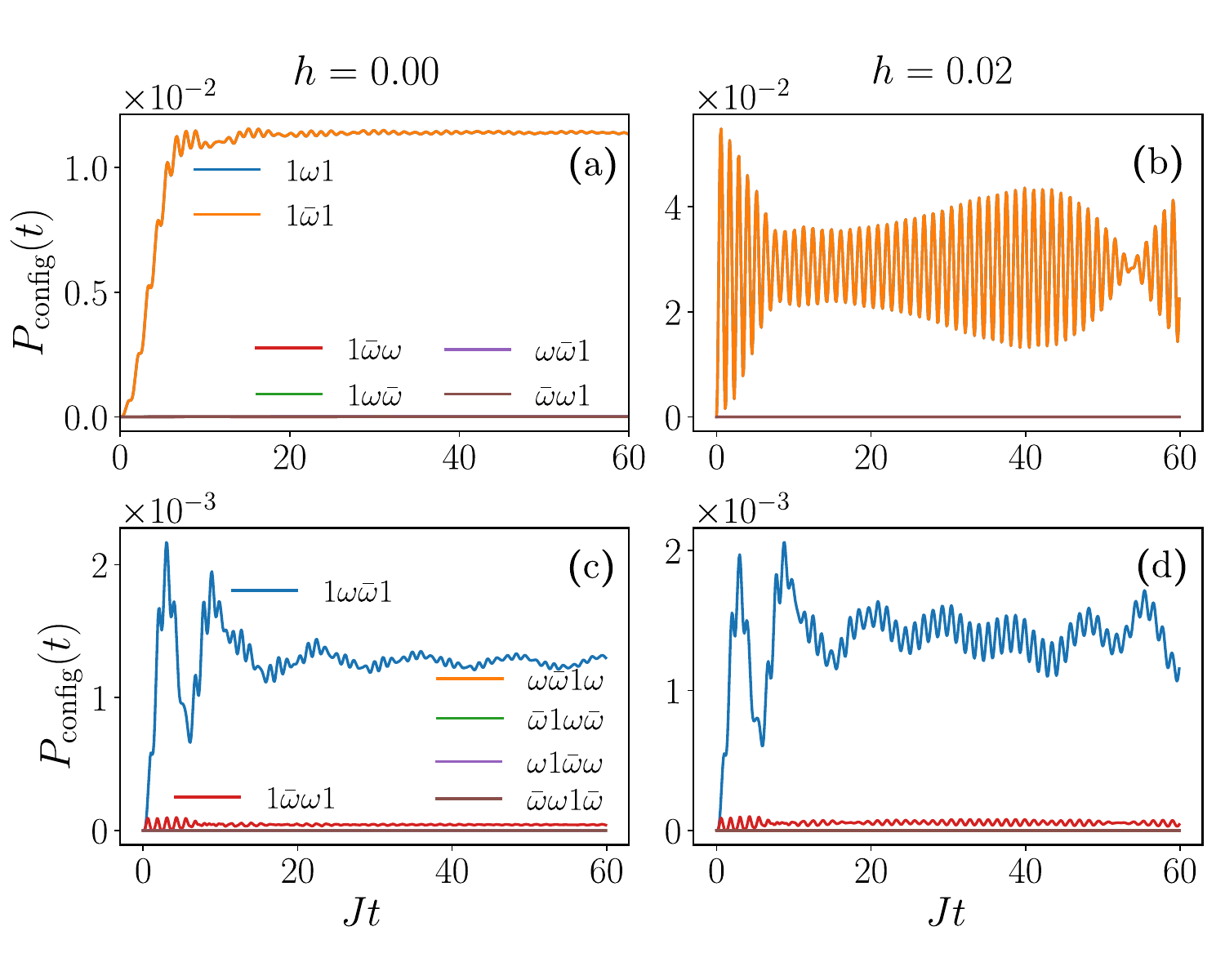}
\caption{
Panel a shows the total probability (over all positions) of two domain wall states corresponding to the ones in Fig \ref{fig:NCDomainWallProb}a after a quench to a Hamiltonian with $\theta=\pi/8$ and $f=0.1$. Unlike the non-chiral case, the domain walls do not scatter into other forms at the boundary resulting in a steady probability.
Panel b shows the same but with a final Hamiltonian that has an additional longitudinal field $h$. Panels c and d shows the probability weight of three domain wall states of type $ABCA$.  
Unlike the non-chiral case, the probabilities of $1\omega \bar{\omega}1$ and $1\bar{\omega}\omega1$ states occur with different probabilities. Note that in Panels a and b lines corresponding to $1\bar{\omega}\omega$,$1\omega\bar{\omega}$, $\bar{\omega}\omega 1$ and $\omega\bar{\omega}1$ overlap on each other. Same is true of $1\omega 1$ and $1\bar{\omega} 1$ lines. In panels c and d lines corresponding to $\omega\bar{\omega}1 \omega$, $\bar{\omega}\omega1 \omega$, $\omega1\bar{\omega} \omega$ and $\omega1 \omega\bar{\omega}$ overlap with each other.
System size is $L=50$. 
\label{fig:C_domainwallprob}}
\end{figure}

\section{Numerical results: Quench into the chiral Hamiltonian}
\label{sec:numResultsC}
Now we focus on dynamics in the system after a quench from the initial, fully ordered state, into the chiral Hamiltonian with finite $f$ and $\theta$. As mentioned in Sec \ref{sec:model}, we focus on $\theta<\pi/6$, where the classical ground state ({\emph {i.e.}} the Hamiltonian ignoring the transverse field) is ferromagnetic and the domain walls are well defined. The main effect of chirality then is to induce different energies to the domain walls of opposite chirality. 

We will begin with a discussion of the probabilities of the domain wall flavors generated after the quench. Figure \ref{fig:C_domainwallprob}a presents the total probabilities of two domain wall states similar to the Fig \ref{fig:NCDomainWallProb}a. Only the $1\omega1$ and $1\bar{\omega}1$ domain walls are generated and these two occur with equal probabilities. The locality of the Hamiltonian does not allow for the formation of $ABC$ domain wall pairs in the bulk.
The post quench Hamiltonian considered here has $\theta=\pi/8$ and $f=0.1$. Using the results at the end of Sec \ref{sec:model}, we see that the gap between the domain wall bands (between the bottom of the upper domain wall band and the top of the lower domain wall band) is $2\sqrt{3}J\sin\theta-4f>0$ and therefore the domain walls bounce back from the boundary without change in its flavor. As a result the total probability of the $1\omega1$ and $1\bar{\omega}1$ domain walls remain steady as seen in Fig \ref{fig:C_domainwallprob}a. This is unlike the non-chiral model discussed previously (Fig \ref{fig:NCDomainWallProb}a). 

Figure \ref{fig:C_domainwallprob}c shows the probabilities of $ABCA$ type three domain wall states. The $1\omega\bar{\omega}1$ states are generated with higher probability than the opposite chirality $1\bar{\omega}\omega1$ type domain walls which has a higher energy. As discussed in Sec \ref{sec:model}, as $\theta$ approaches $pi/6$, the energy of the $1\bar{\omega}$ domain wall becomes close to that of a pair of domain walls $1{\omega}\bar{\omega}$. As a result two domain walls can evolve into three domain wall states. Numerics show that the three domain walls proliferate as $\theta\to\pi/6$. We will leave the analysis of this regime for later studies.

\begin{figure}
\includegraphics[width=\columnwidth]{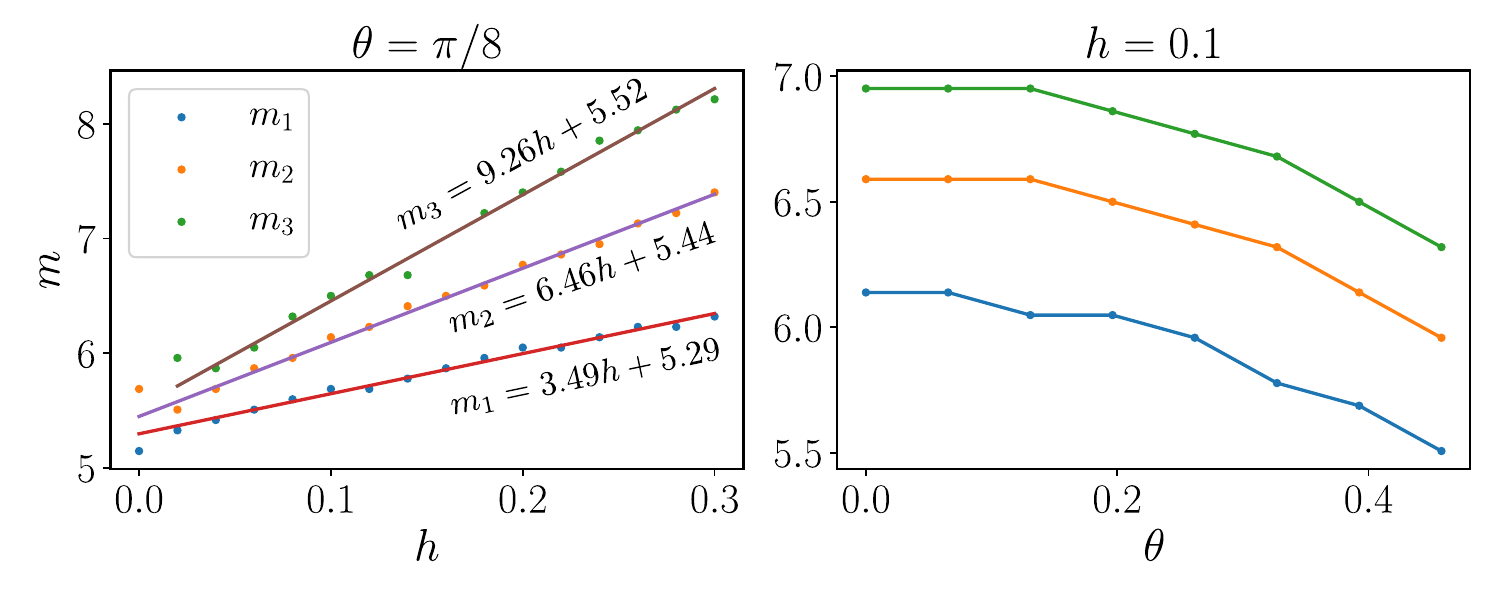}
\caption{
Energy of bound domain wall pairs extracted from the power spectrum of magnetization in the same manner as in Fig \ref{fig:Mag_NC}. Panel a shows the masses as a function of the longitudinal field for a fixed $\theta$. Panel b shows the dependence of the masses on $\theta$ for fixed $h$. Spectra are calculated from magnetization time series upto time $t=90$ in a system of size $L=40$. The scatter in the data is primarily caused by the finite frequency resolution in a Fourier transformation of data over a finite range of time.
\label{fig:C_mass}}
\end{figure}

\subsection{Magnetization}
As in the case of the non-chiral model, magnetization decays linearly with time at short times (Fig \ref{fig:Mag_NC}a) with a small oscillatory component of (angular) frequency given by the total mass of a pair of opposite chirality domain walls, namely 
\begin{equation}
m_2(\theta)=\epsilon_{k=0}^-+\epsilon_{k=0}^+=6J\cos\theta - 4f.
\end{equation}
Upon adding a longitudinal field, bound domain wall pairs are formed whose masses can be inferred from the magnetization oscillations as described in the Sec \ref{sec:numResultsNC}. Fig \ref{fig:C_mass} summarizes the dependence of the masses on $\theta$ and $h$; masses appear to increase linearly with $h$ and decrease monotonically with $\theta$ in the ranges considered.

\begin{figure}
\includegraphics[width=\columnwidth]{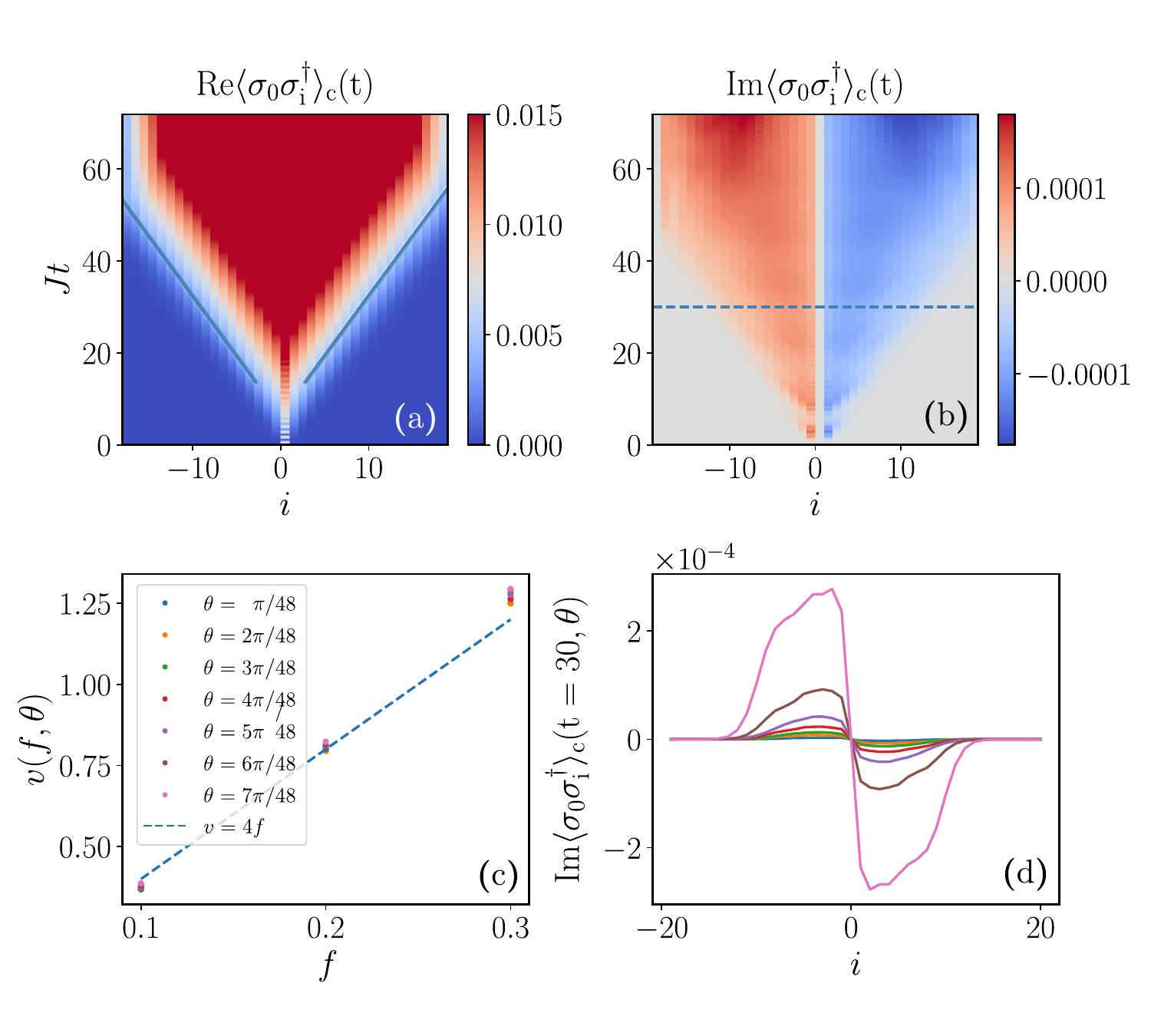}
\caption{Panels a and b show the real and imaginary parts of the correlator $C(i,t)$ after the fully ordered state is quenched to a final Hamiltonian with $f=0.1, \theta=\pi/8$. Straight lines overlayed in the figure showing the rate of spread of correlations are obtained by fitting to constant $C$ contours. The slope of the line is consistent with the expected rate of spread of correlations $2v_g=4f$.
Panel c shows the rate of spread as a function of $f$ for different $\theta$ values. The dotted line shows the expected dependence $4f$.
In Panel d, ${\rm Im}\langle\sigma_0\sigma_i^\dagger\rangle_c(t)$ as a function of position is shown for different $\theta$ values and a fixed time slice $t$ (corresponding to the time slice indicated by the horizontal line in panel b).
System size used for the calculation is $L=40$.
\label{fig:C_2ptCorrelations}}
\end{figure}

\begin{figure*}
\includegraphics[width=\textwidth]{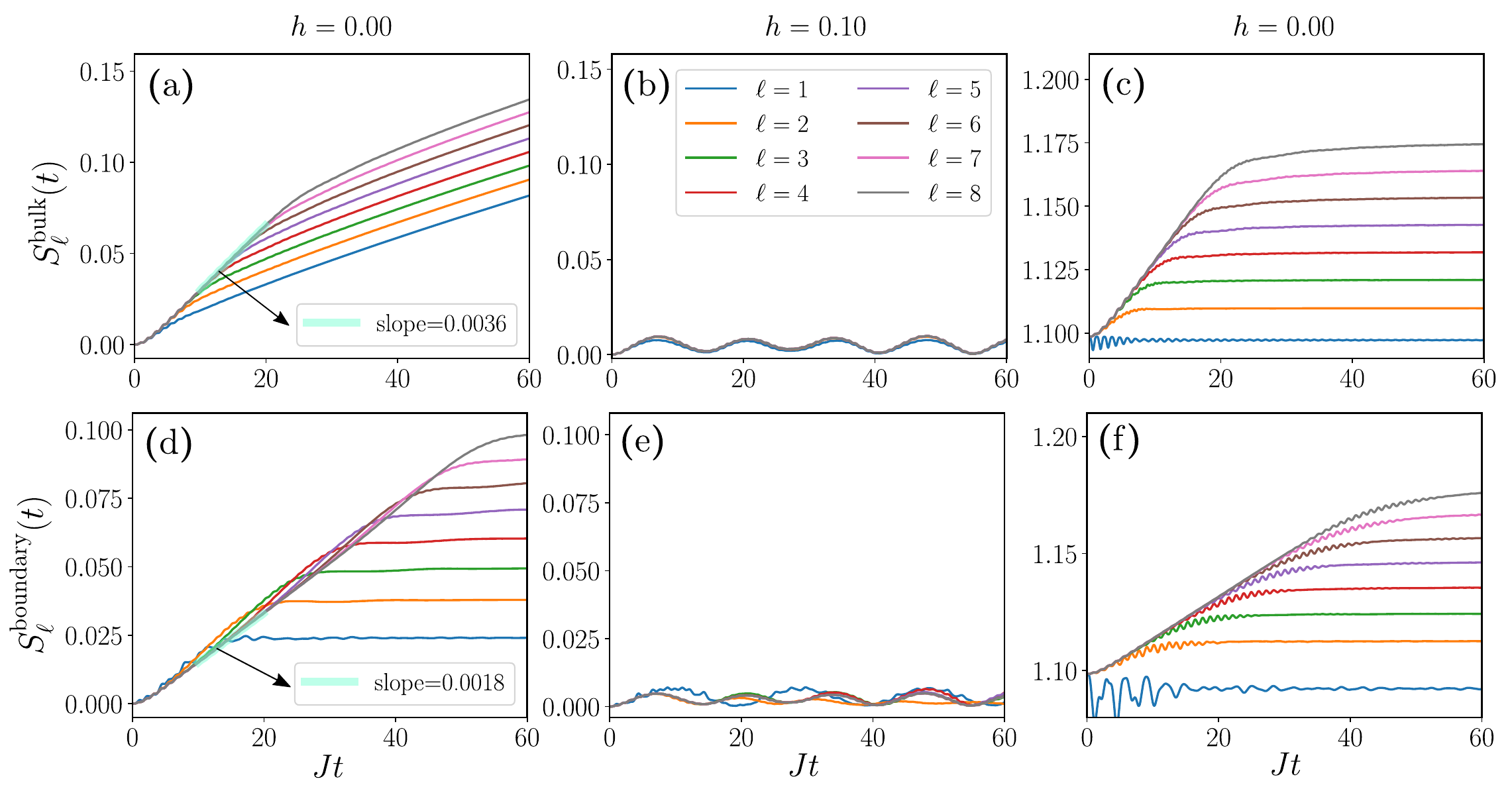}
\caption{
Panels a and d show the entanglement entropy as a function of time for subsystems located in the bulk and at the edge of the system.
Panels b and e show the same for the cases where the final Hamiltonian has an additional longitudinal field. 
Panels c and f show the results for the case where the initial state is a parity eigenstate of the form $|..1111...\rangle + |...\omega\omega\omega...\rangle + |...\bar{\omega}\bar{\omega}\bar{\omega}...\rangle$. 
Above results are obtained in a system of size $L=40$ and for a final Hamiltonian with $f=0.1$ and  $\theta=\pi/8$.
\label{fig:C_entropy}}
\end{figure*}

\subsection{Two point correlations}
The connected two point correlations at equal times $C(r,t)=\langle \psi(t)|\sigma_0 \sigma_r^\dagger|\psi(t) \rangle $ is shown in Fig \ref{fig:C_2ptCorrelations}. Since the domain wall velocities are independent of $\theta$  ($v_g\sim 2f$), the rate of spread of correlations ($2v_g$) remain the same as in the non-chiral model (Fig \ref{fig:C_2ptCorrelations}c).

Spatial parity is not a symmetry of the dynamics, therefore the imaginary part of the correlations is not necessarily zero. An expansion of the $\psi(t)$ in the computational basis ({\emph {i.e.}} eigenbasis of $\sigma$) together with the results in Fig \ref{fig:C_domainwallprob} indicates that the complex part of the correlations (Fig \ref{fig:C_2ptCorrelations}b) arise due to an excess occurrence of three domain wall states of one chirality over the other. 
Since the three domain wall states have low abundance, the imaginary part of the correlations is much smaller than the real part (Fig \ref{fig:C_2ptCorrelations}a,b). The difference between probabilities of opposite chirality three domain wall states increases with $\theta$. This manifests in the increase with $\theta$ of the imaginary part of the correlations (Fig \ref{fig:C_2ptCorrelations}d).

\subsection{Entanglement entropy}
Now we describe the results for entanglement entropy growth after a quench into the chiral Hamiltonian. The entropy of small subsystems in the bulk (Fig \ref{fig:C_entropy}a) grows linearly with time until $t\sim \ell/2v_g$ (where $v_g\sim 2f$). This regime is, as explained in Sec \ref{sec:numResultsC}, described by population of new states with flipped spin domains. Following the saturation of this mechanism, the two oppositely ordered states are populated as the domain walls exit the system, resulting in further growth of the entropy. As in the case of the non-chiral model, when the initial state is a parity eigenstate, the entropy grows linearly  from $\ln 3$ (entropy of subsystems of a parity eigenstate) and saturates at a time $t\sim\ell/2v_g$  (Fig \ref{fig:C_entropy}f). The growth is strongly suppressed in the presence of a longitudinal field (Fig \ref{fig:C_entropy}b).

Entanglement entropy of small subsystems located at the boundary of the system grows linearly with time till $\ell/v_g$ at a rate half that of the subsystems in the bulk. This is shown in Fig \ref{fig:C_entropy}d. 
In the case of the quench to the non-chiral model, the entanglement entropy in the subsystem located at the boundary continues to grow after time $\ell/v_g$. In contrast, here the entanglement entropy saturates to a constant (Fig \ref{fig:C_entropy}d).
This can be understood to arise from scattering properties at the boundary. In the chiral case, the domain walls of the form $AB$ that reach the right hand side boundary are reflected back as a domain wall of the type $AB$.
When the domain walls exit the subsystem, they leave the subsystem in the same state as the initial state $|...1111...\rangle_A$. 
There is no increment in the population of the oppositely ordered states. This is unlike the non-chiral model.

\begin{figure}
\includegraphics[width=0.8\columnwidth]{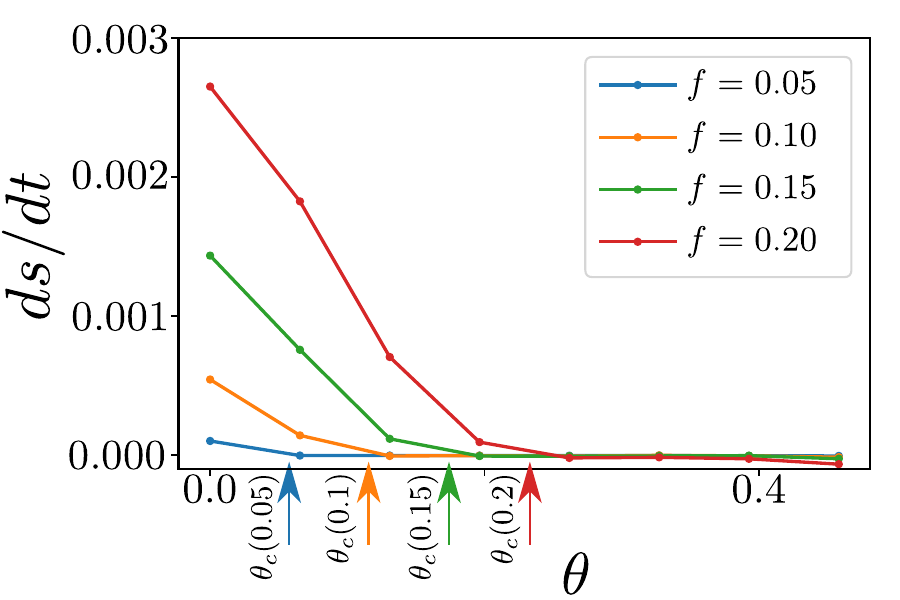}
\caption{
Rate of change of entropy in a subsystem located at the edge at an instant ($Jt=60$) after the saturation time $\ell/v_g$ is shown as a function of $\theta$ for different $f$ in the final Hamiltonian.
The arrows are crossover $\theta_c$ estimated for each $f$ based on Eqn \ref{eq:theta_c}. Entropy saturates for $\theta>\theta_c$.
\label{fig:dsdt}}
\end{figure}

In the non-chiral case, the incoming $AB$ domain wall reflects at the open boundary as an $AC$ domain wall. $AB\to AB$ type scattering (as opposed to $AB\to AC$) occurs if the opposite chirality domain walls have bands (Sec \ref{sec:model} and Fig \ref{fig:BoundaryReflection}b,c) that do not overlap {\emph {i.e.}} if 
\begin{equation}
2\sqrt{3}J\sin\theta>4f.\label{eq:theta_c}
\end{equation}
This is verified in Fig. \ref{fig:dsdt} which shows the rate of change of entropy  after the expected saturation time $\ell/v_g$ in the subsystems located at the system edge, plotted as a function of $\theta$. The rate of change is $0$ for large $\theta$ and non-zero at small $\theta$ with an $f$-dependent crossover $\theta_c$ that is consistent with the above estimate ($\theta_c(f)\sim \sin^{-1}\frac{2f}{\sqrt{3}J}$, marked in the figure with arrows).

\section{Summary and Conclusion}
\label{sec:conclusions}

In this work, we have explored post quench domain wall dynamics in the ferromagnetic chiral clock model. Using finite size simulations, we have addressed the evolution of magnetization expectation values, equal time two point correlation functions, and entanglement growth, and a microscopic picture based on effective dynamics of single domain walls has been presented.

Entanglement growth and spread of correlation happen through evolution of domain-wall-pair states. Irrespective of  $\theta$, domain-wall-pair states of the type $1\omega1$ and $1\bar{\omega}1$ form with equal probability from all points in the system immediately after the quench. Domain walls propagate with a maximal group velocity $v_g = 2f$ independent of the chirality parameter $\theta$. As a consequence there is no qualitative difference between the non-chiral and chiral model in the entanglement and correlation spread in the bulk. In the non-chiral model, total probability of $1\omega1$ and $1\bar{\omega}1$ states decay linearly with time as the domain walls scatter at the boundary and convert to $1\omega\bar{\omega}$, $1\bar{\omega}{\omega}$ due to collisions with the right boundary and to $\bar{\omega}{\omega}1$ and ${\omega}\bar{\omega}1$ due to collisions on the left. In the chiral model, there is no such scattering to different domain wall types. Three-domain-wall states of the form $1\omega\bar{\omega}1$ and $1\bar{\omega}\omega 1 $ are also generated with smaller probabilities compared to two-domain-wall states. In the chiral model the two types of three-domain-wall states are generated with unequal probabilities. 

Magnetization decays linearly with time at short times accessible within our simulations. Oscillations around the linear decay have a frequency equal to the energy cost of two domain walls namely $6J\cos\theta-4f$. In the presence of a longitudinal field that couples to $\sigma+\sigma^\dagger$, domain wall pairs form bound states of energies that appear to increase linearly with the field and decrease with the chirality.

Equal time two point correlations spread with the same speed $2v_g\sim 4f$ in both the chiral and non-chiral models. Imaginary part of the specific correlation $\langle \sigma_0(t)\sigma_r(t)^\dagger \rangle_c $ reflects the relative abundances of the opposite chirality three-domain-wall states. It is zero for the non-chiral model and increases  in magnitude with $\theta$. 

Entanglement entropy in subsystems located in the bulk shows a linear growth, and saturates at a characteristic time scale $\tau_s\approx \ell/2v_g$. In small subsystems located in the bulk, a subleading growth of entanglement is seen after this time. In the non-chiral model, the similar behavior is seen even in the subsystems located at the boundary of the system (till a time $\tau_s~\ell/v_g$). In the chiral models, with the chirality parameter $\theta > \sin^{-1}\frac{2f}{\sqrt{3}}$, the entanglement saturates to a constant. 

We find that a linear-in-time entanglement growth is seen even outside the ferromagnetic regime of $\theta$ that we have studied. However, a simple isolated domain wall description is not sufficient to understand the behavior. At larger values of $\theta$ above $\pi/3$ where the ground state is not ferromagnetic, chirality in the ground state magnetization will have a more complex interplay with a longitudinal field than in the small $\theta$ cases we have studied. The second parameter in the model ($\phi$) will act as an effective magnetic field to the domain wall particles, bringing in richer structures in the quenches in the model. We leave the exploration of the dynamics in the extended parameter space of the model for future studies.

\begin{figure}
\includegraphics[width=\columnwidth]{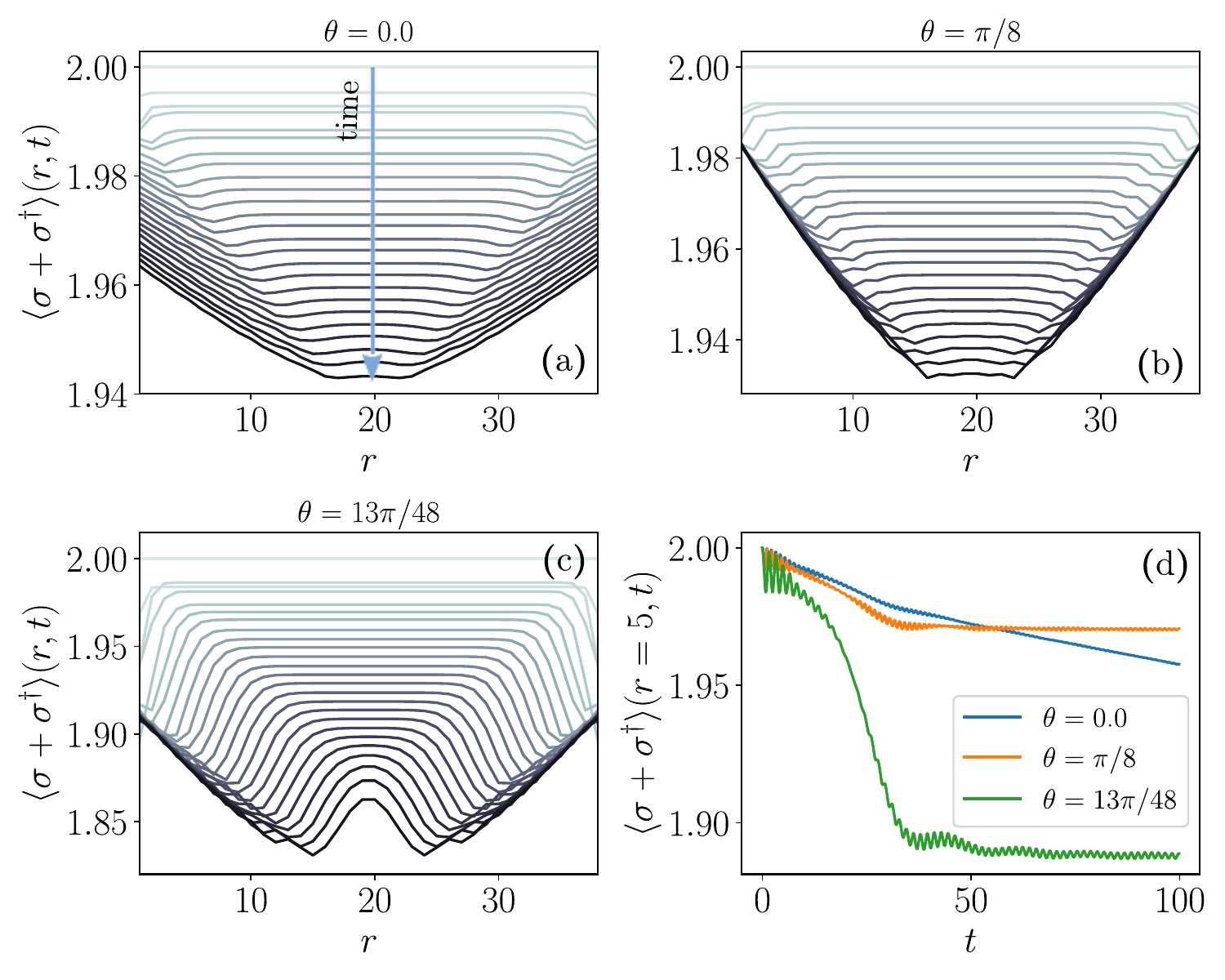}
\caption{
Magnetization profile in the chain with open boundary conditions at the left and right ends. The magnetization at the boundary of the chain decays with time in the non-chiral model (panel a), whereas it saturates to constant in the chiral model (panels b,c). The saturation values vary linearly with distance from the boundary. Panel d shows the magnetization at a fixed position near the edge as a function of time for three different values of $\theta$.
\label{fig:edgeMagnetization}}
\end{figure}

The saturation of entanglement in the small subsystems at the edge of the system points to the inability of the spreading domains to thermalize the spins at the boundary of the system into an equally probable mixture of $1$, $\omega$ and $\bar{\omega}$. As a consequence the initial magnetization survives at long times after quench. Careful accounting of the domain walls at the boundary after the saturation time indicates that the density of flipped spins near the boundary linearly changes with distance from the boundary. Consequently the post quench magnetization in the chiral model shows a linear decay of magnetization away from the boundary (Fig \ref{fig:edgeMagnetization}).

In contrast, the non-chiral model shows a magnetization that appears to decay to $0$ at the boundary. 
Thus the $\theta$ dependent boundary scattering presents a peculiar scenario of a non thermal steady state near the boundary of this chain. Such a mechanism for failure of thermalization is related to the long coherence times of boundary spins in models carrying boundary zero modes in Jordan Wigner transformed dual description.\cite{Kemp_2017,PhysRevX.7.041062}

\acknowledgments
Calculations were performed using codes built on ITensor Library\cite{itensor} and verified in smaller systems with Petsc/Slepc based codes. SGJ acknowledges DST/SERB grant ECR/2018/001781, IISER-CNRS joint grant, and National Supercomputing Mission (Param Brahma, IISER Pune) for computational resources and support.

\bibliography{Clock}
\appendix

\section{Fourth order accurate time evolution}
\begin{figure}
\includegraphics[width=0.8\columnwidth]{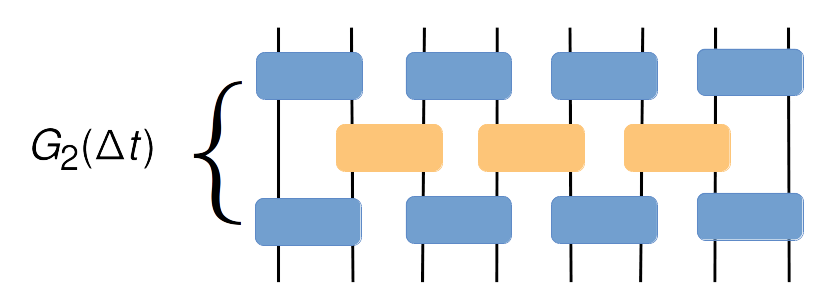}
\caption{MPO sequence for second order approximant $G_2(x)$ where blue and yellow colored two site MPO corresponds to $e^{\imath A_n x/2}$ and $e^{\imath B_nx}$ respectively.\label{fig:MPOsequence}}
\end{figure}

Here we summarize the fourth order approximant\cite{Hatano2005} for the exponential of a time independent local Hamiltonian, which is used to construct the unitary time evolution operator. Any Hamiltonian on a chain with only nearest neighbor couplings can always be split into two parts $\mathcal{H}_{\rm o}$ and $\mathcal{H}_{\rm e}$ acting only on odd and even bonds respectively. The two parts $\mathcal{H}_{\rm o}$ and $\mathcal{H}_{\rm e}$ can be further written as -
\begin{equation}
\mathcal{H}_{\rm o}=\sum_n A_n\text{ and }\mathcal{H}_{\rm e}=\sum_n B_n
\end{equation}
where $A_n$ for $n=1,2,3..$ has support on sites $2n-1$ and $2n$ only, and $B_n$ has support on sites $2n$ and $2n+1$ only. Since $A_n$s commute with each other and $B_n$s commute with each other, exponential of the $-\imath \mathcal{H}_{\rm o} dt$ and $-\imath\mathcal{H}_{\rm e} dt $ can be written as product of two site operators $\prod_n e^{-\imath dt A_n}$ and $\prod_n e^{-\imath dt B_n}$ respectively and these can be efficiently implemented as matrix product operators. Note that $A_n$ and $B_n$ do not generically commute with each other if their supports overlap and the full unitary is not the product of these two exponentials. However by suitable combination of such terms, the exponential of the Hamiltonian can be written to an arbitrary finite order of accuracy in $dt$ using a fractal decomposition where a higher order approximant is obtained recursively from lower order approximants.\cite{Hatano2005}. The fourth order approximant $G_4$ of the unitary operator $\exp(-\imath dt \mathcal{H})$ used in our calculation is given by 
\begin{equation}
G_4(dt)=G_2(s_2\,dt)^2 G_2((1-4s_2)\,dt) G_2(s_2\,dt)^2
\end{equation}
where $s_2=1/(4-\sqrt[3]{4})$ and $G_2$ is second order approximant which is given by
\begin{equation}
G_2(dt)=e^{-\imath\mathcal{H}_{\rm o}\,dt/2}e^{-\imath\mathcal{H}_{\rm e}\,dt}e^{-\imath\mathcal{H}_{\rm o}\,dt/2}
\end{equation}
Using the commutative properties operators of $A$ and $B$, $G_2(dt)$ can be represented by following the MPO sequence shown in Fig \ref{fig:MPOsequence}

Each two site MPO in Fig \ref{fig:MPOsequence} is of the form $e^{Mdt}$ and can be approximated as $E=\sum_{s=0}^4 M^s \frac{dt^s}{s!}$. $E$ obtained  numerically can be expanded as $\sum_{ij}\lambda_{ij} O_i\otimes O_j$ 
where $O_i\in\{\mathbb{I},\sigma,\sigma^\dagger,\tau,\tau^\dagger,\sigma\tau,\sigma\tau^\dagger,\sigma^\dagger\tau,\sigma^\dagger\tau^\dagger\}$ and $\lambda_{i,j}=\text{tr}(EO_i^\dagger\otimes O_j^\dagger)/9$.



\end{document}